\documentclass[mathleft
]{an}
\usepackage{graphicx}
\usepackage{rotating}
\usepackage{lscape}
\usepackage{longtable}
\usepackage{times}
\begin{document}

\sloppy

\Pagespan{789}{}
\Yearpublication{2011}%
\Yearsubmission{2010}%
\Month{11}%
\Volume{999}%
\Issue{88}%

\title{A review of East Asian reports of aurorae and comets circa AD 775}

\author{J. Chapman\inst{1} \thanks{E-mail: jessechapman@berkeley.edu}, D.L. Neuh\"auser\inst{2}, R. Neuh\"auser\inst{3}, M. Csikszentmihalyi\inst{1}}

\titlerunning{Chinese aurorae in the AD 770s}
\authorrunning{Chapman et al.}

\institute{
$^{1}$ Department of East Asian Languages and Cultures, UC Berkeley, Berkeley CA, 94720, United States \\
$^{2}$ Schillbachstrasse 42, 07745 Jena, Germany \\
$^{3}$ Astrophysikalisches Institut, Universit\"at Jena, Schillerg\"asschen 2-3, 07745 Jena, Germany
}

\received{2015 March 4}
\accepted{2015 June 4}
\publonline{2015 Aug 1}


\abstract{Given that a strong $^{14}$C variation in AD 775 has recently been suggested 
to be due to the largest solar flare ever recorded in history, 
it is relevant to investigate whether celestial events observed around that time 
may have been aurorae, possibly even very strong aurorae, 
or otherwise related to the $^{14}$C variation 
(e.g. a suggested comet impact with Earth's atmosphere). 
We critically review several celestial observations from AD 757 to the end of the 770s,
most of which were previously considered to be true, and in some cases, strong aurorae;
we discuss in detail the East Asian records and their wording.
We conclude that probably none among the events 
after AD 770
was actually an aurora,
including the event in AD 776 Jan, which was misdated for AD 774 or 775;
the observed white {\em qi} phenomenon that happened {\em above the moon} in the south-east
was most probably a halo effect near the full moon 
-- too late in any case to be related to the $^{14}$C variation in AD 774/5.
There is another report of a similar (or identical) white {\em qi} phenomenon {\em above the moon},
reported just before a comet observation and dated to AD 776 Jan;
the reported comet observed by the Chinese was misdated to AD 776, but actually sighted in AD 767.
Our critical review of East Asian reports of aurorae circa AD 775
shows some very likely true Chinese auroral displays observed 
and reported for AD 762; there were also several events 
prior to AD 771 that may have been aurorae but are questionable.
}

\maketitle

\section{Introduction: $^{14}$C variation AD 774/5}

A strong variation in $^{14}$C data with a 1-2 yr time resolution around AD 775 was detected in two Japanese 
(Miyake et al. 2012), one German (Usoskin et al. 2013), one American
and one Siberian tree (Jull et al. 2014). 
A sudden increase in $^{14}$C could indicate a strong input of highly energetic particles 
or $\gamma$-rays into the Earth's atmosphere. 
A supernova has almost been excluded as the cause of the event (Miyake et al. 2012, Hambaryan \& Neuh\"auser 2013); 
both a solar super-flare (e.g. Melott \& Thomas 2012, Usoskin et al. 2013) and a Galactic short gamma-ray burst 
(Hambaryan \& Neuh\"auser 2013) have been proposed as alternatives. 
While such a large solar super-flare has (otherwise) never been recorded in the last 3000 
(M12) or even 11,000 yr (Usoskin \& Kovaltsov 2012) in Intcal $^{14}$C data (rate possibly being zero), 
the rate of short gamma-ray bursts is known from observations of other galaxies 
to be non-zero, but possibly very low. 
The super-flare hypothesis was also considered unlikely by Cliver et al. (2014) 
and Neuh\"auser \& Hambaryan (2014) by comparison with the strongest flares 
of the last two centuries. Hence, it is relevant to search for alternative suggestions.

Liu et al. (2014) pointed to an impact of a comet with Earth's atmosphere on AD 773 Jan 17, 
presumably seen in a $^{14}$C peak in corals at that time, which was, however, 
rejected for three reasons: As mentioned in Neuh\"auser \& Hambaryan (2014), 
given the carbon cycle, $^{14}$C would first be incorporated by trees, 
and later by corals, so that the sequence of events suggested by Liu et al. (2014) is not possible. 
Both Usoskin \& Kovaltsov (2014) and Melott (2014) showed 
that comets could not have delivered as much $^{14}$C as needed for the AD 775 variation. 
Then, Chapman et al. (2014) clarified that the Chinese comet observed AD 773 Jan 17 
was a normal comet with a long tail, which was then also observed in Japan three days later, 
and thus could not have collided with the Earth on Jan 17.

Usoskin et al. (2013) claimed to have found {\em a distinct cluster of aurorae between AD 770 and 776}
which suggests {\em a high solar activity level around AD 775}.
They list 14-16 different events as aurorae from AD 765 to 786, 
including some presumable aurorae observed in China from AD 770 to 775. 
More recently, Zhou et al. (2014) claimed to have found worldwide super-aurorae 
connected with the suggested solar super-flare: 
\begin{quotation}
The strongest AD 775 auroras in the past 11400 years were first successfully 
identified with the historical records ..., the super auroras were generated 
in Jan AD 775 and lasted about 8h.
\end{quotation}
This claim is based mainly on one alleged aurora report from China (but see our Sect. 2.7), 
but also on two doubtful events from Europe, 
which were both misdated (Neuh\"auser \& Neuh\"auser 2014, 2015).

We review all previously suggested aurora reports as observed from East Asia from AD 757 to 779.
Most of the known East Asian aurora records are found in Matsushita (1956), 
Keimatsu (1973, 1974),\footnote{From Matsushita (1956) and
Keimatsu (1973, 1974), we include only those which were classified by them with
(a high) probability 1-3 (out of five); in Keimatsu (1973, 1974), this meant {\em certain} for 1,
{\em very probable} for 2, and {\em probable} for 3; events classified by
Keimatsu (1973, 1974) with lower probability are discussed in this paper only, because
they were classified as true aurorae by others, such as Yau et al. (1995) or Xu et al. (2000).}
Dai \& Chen (1980), Yau et al. (1995), or Xu et al. (2000).
In Sect. 2, we discuss the events which obviously were not aurorae,
in Sect. 3, list likely true aurorae,
and in Sect. 4, we also give those events, which may or may not have been aurorae.
In addition, we also review 
East Asian
observations of comets reported for the mid AD 770s,
one of them being in connection with an appearance of white {\em qi}
(Sect. 5). 
We conclude our findings in the last section.

First, we comment briefly on medieval Chinese astronomy (and astrology) and 
in particular their knowledge of what we now understand as aurorae.

Any study of aurorae in medieval China is complicated by the fact 
that there was no discrete concept of aurorae as such in medieval Chinese astronomy.  
Scholars have variously identified observations 
of flowing stars ({\em liu xing}) or stars that fall ({\em xing yun}), 
various sorts of halos ({\em huan}), and {\em qi} as aurorae.  
Flowing stars and stars that fall in almost every case should be identified with meteors
or bolides, while {\em huan} ({\em rings}) are most likely lunar or solar halo displays;
flowing stars in some instances may also refer to comets.

The most likely instances of aurorae in Chinese historical records are identified as {\em qi}, 
yet not all or even most observations of {\em qi} were indeed aurorae.  
While {\em qi} is variously translated as {\em ether(s)} or {\em vapour(s)}, 
material objects, including clouds, planets, stars, comets, and meteors, 
were thought to be constituted of {\em qi} in Chinese cosmology.  
Because {\em qi} was thought to emanate from the Earth itself, 
often in response to developments in the politico-religious sphere of the imperial court, 
explanations for aberrant astronomical and meteorological phenomena were grounded in politics.  

In the first major treatise on astro-omenology, 
the {\em Tianguan shu} (Treatise on the Celestial Office) Director of Astronomy/Senior Archivist 
Sima Qian (ca. BC 145 to ca. 86) wrote in the late 2nd century BC: 
\begin{quotation}
Heaven has the sun and moon, and Earth has yin and yang.  
Heaven has the Five Planets and Earth has the Five Resources.  
Heaven has its arrayed lodges, and Earth has its regions.  
The Three Luminaries [i.e. the sun, moon, and stars] are the essence of yin and yang.  
Qi originates in the Earth, and the sages unify it and put it to order. 
(Shiji 27.1342 / Sima Qian 1959)
\end{quotation}

Two centuries later, circa AD 100, Ban Zhao, in her {\em Treatise on Celestial Patterns}, 
explained virtually all aberrant celestial phenomena, including eclipses, halos, strange clouds, and abberations
of {\em qi} as issuing from ritual or administrative failures on the part of the ruler and his court: 
\begin{quotation}
Portents originate in the earth and erupt above into the heavens.  
When the administration fails here, then transformations appear there, just as shadows are signs of their form, 
and echoes are responses to sounds.  
This is why the clear-sighted ruler sees them and awakens, 
putting himself in order and rectifying his affairs.
(Hanshu 26.1273 / Ban Gu et al. 1962)
\end{quotation}

Aurorae were no exception. Two events listed as probable true aurorae in sections 3.1 and 3.2 below, for example, 
are immediately followed by a description of political events that occurred in the following year.  
The {\em Jiu Tang shu} ``Treatise on Celestial Patterns''
associates the appearance of red auroral lights in the sky with the chaotic state of the empire:  
\begin{quotation}
In the tenth month of the following year, the Tufan took Chang'an, 
and Emperor Daizong graced Shaanzhou with his presence as he fled from the northern tribes.
(Jiu Tang shu 36.1325)
\end{quotation}
We are to understand aurorae, and aberrant astronomical and meteorological phenomena in general, as portents, 
the meanings of which become clear through the writing and reading of history.

Standard historical treatises on celestial patterns ({\em tianwen}) often refer to political events 
happening at court which explain the meaning of celestial signs such as aurorae, 
thus offering an implicit causal explanation of such events in astrological terms.  
Unlike Ptolemaic astrology, human beings in the Chinese system were not passive subjects of astrological forces, 
but their ritual, political, and military actions were thought to produce astronomic and meteorological phenomena.  
For this reason, court astronomers carefully chronicled aberrant and/or transient astronomical and meteorological phenomena, 
laying the basis for the compilation of the treatises on celestial patterns in successive dynasties.  

The gap in time between when the observations were originally recorded and when they were ultimately included 
in historical treatises creates another problem.  
Though the Chinese meticulously recorded the time when observations occurred based on 
a clear lunar calendar and twelve 
daily two-hour periods or double-hours,
printing had not yet developed in the Tang dynasty (AD 618-907) and so few copies of the records could be kept.  
Mistranscription and material decay inevita\-bly introduced errors into the texts.  

Time-keeping for astronomical events was often quite precise, 
recording the reign year, month, day, and double-hour when a given event occurred.  
In contrast to the five night-watches, which varied in length at different times of year,
the twelve double-hours (for 24h per day) were equal in length (120 min each), measured by clepsydra (water clocks).  
The mid-point of the first double-hour, {\em zi}, was reckoned as midnight (Wilkinson 2000).
The first night-watch started with the end of the evening twilight (dusk), 
i.e. 36 min after sunset (since AD 25),
and the fifth night-watch ended with the start of the morning twilight (dawn), 
i.e. 36 min before sunrise (Stephenson 1997).
For the convenience of the reader, we follow the standard practice of converting Chinese dates
to the Julian calendar. When neccessary, we also convert the five night-watches to our 
clock-hours (in local time) for the given location and time 
(taking into account the 4-day difference between
the Julian and Gregorian calendar by the time of the 8th century).

In terms of practical observation of celestial patterns (i.e. {\em tianwen} -- a term often imprecisely translated 
as astronomy or astrology), 
clouds ({\em yun}) and {\em qi} 
were included in a single category, {\em yunqi}, and recorded 
in the same section of the astronomical treatises in the two standard histories of the Tang dynasty.
There are numerous astronomical reports for the period under discussion, AD 757-779; 
the lack of certain kinds of astronomical events in the historical record is not due to a mere lack of reports.

Most recently, Stephenson (2015) compiled a list of East Asian (and European) aurorae for the period AD
767 to 779 in the {\em Jiu Tangshu}. For this period, he presents seven texts,
six of which he lists as possible aurorae, we also address these six texts. 
The remaining text concerns {\em black vapour} ({\em heiqi})
(AD 775 Oct 16, similar to an entry for AD 786 Jan 21), 
which neither we, nor any previous scholars, consider to have been an aurora.  
Stephenson (2015) also does not classify the {\em black vapour} as an aurora, 
but as dark clouds seen against a bright (halo) background. 
His translations of the six common texts are laregly consistent
with ours. We comment on differences in Sects. 2.7 and 6.

\section{East Asian observations misinterpreted as aurorae from AD 757 to 779}

We will first discuss a few presumable aurora observations between AD 757 and 779,
that have been misinterpreted as aurorae.
The original Chinese texts are given in Fig. 1.
We give our translations, sometimes compared to previous translations.

\begin{figure*}
{\includegraphics[angle=0,width=17cm]{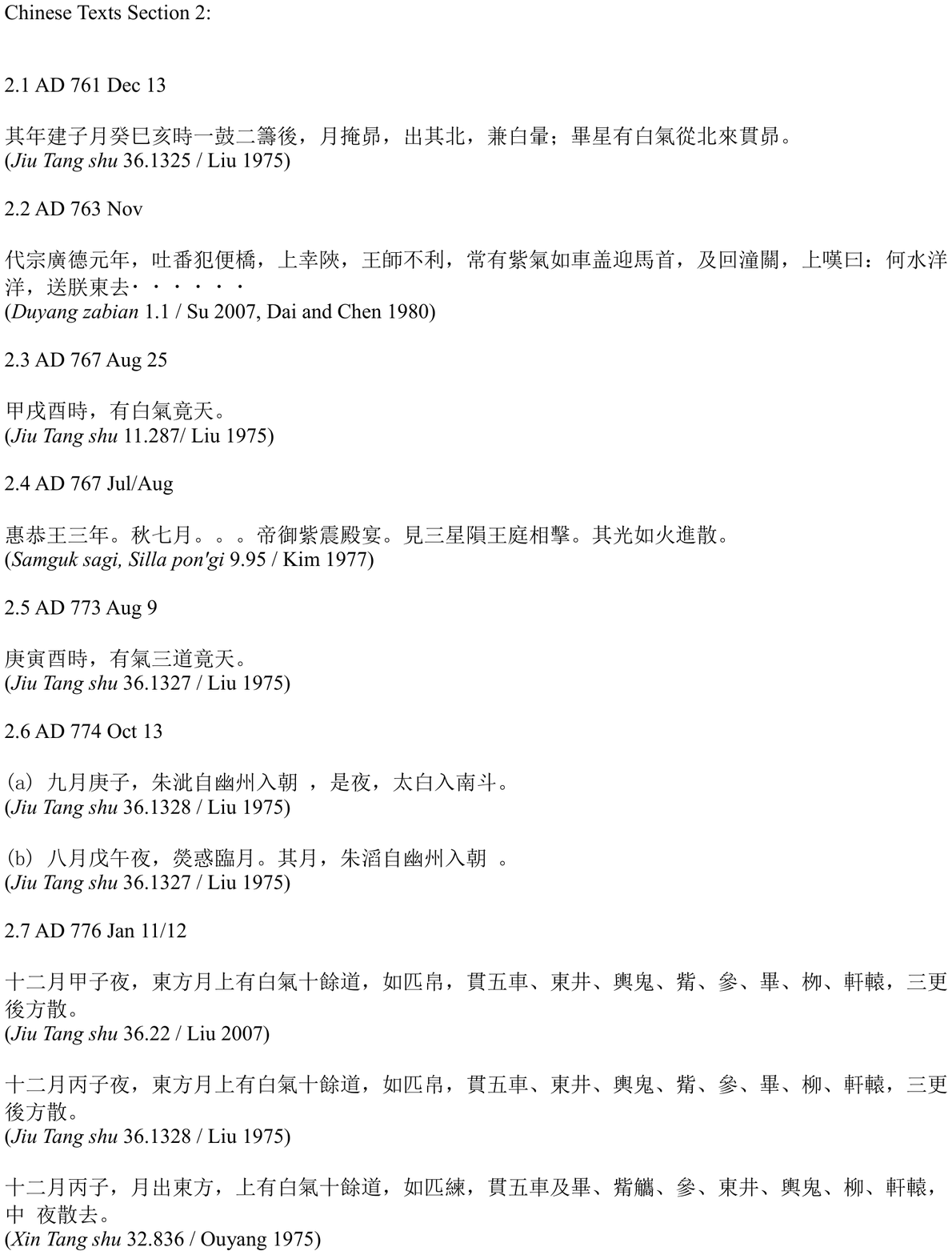}}
\caption{Here, we show the Chinese texts related to events misinterpreted as aurorae. 
Translation and discussion are given in the Sect. 2.}
\end{figure*}

\subsection{AD 761 Dec 13: {\em the moon covered Mao}}

For AD 761 Dec 13, we find in Liu Xu's (AD 887-946) {\em Jiu Tang shu} 36.1325 
(Old History of the Tang Dynasty) for the Tang capital Chang'an,
which corresponds to the modern city of Xi'an, Shaanxi province, China
(square brackets by us): 
\begin{quotation}
In that year [the second year of the Dayuan reign period, AD 761], 
on the guiyi day of the 11th month, 
at the hai double-hour [21-23h] 
after the first night drum [in] the second fifth-hour [i.e. in the 2nd fifth of the double-hour]  
the moon covered Mao [the Pleiades], and then emerged to its north.  
It was surrounded by a white halo. The stars of Bi [in Tau] had white qi amongst them which 
followed the moon north to penetrate Mao.
\end{quotation}

The {\em hai} double-hour (one of 12 double-hours measured by water clocks) can be considered
as quite exactly from 21 to 23h local time; the {\em first night drum} is the border between the first and
second of five night-watches;\footnote{The night was devided into five night-watches of equal lengths, 
which had different lengths throughout the year (like sun-dial hours).} 
for Xi'an, China ($34^{\circ}16^{\prime}$N $108^{\circ}54^{\prime}$E, 7.26h east of Greenwich),
on AD 761 Dec 13, the second night-watch ran from about 20:03h to 22:38h,
and the event is specified to have occurred during the second fifth of the night watch, 
from about 20:34-21:05h. Hence the event took place close to 21h local time.
Indeed, an occultation of the Pleiades (Mao) by the moon in the SE occurred at that time.
Immediately after the occultation, the moon was slightly north relative to the Pleiades,
as reported. In addition, the moon was said to be {\em surrounded by a white halo}, i.e., a
lunar halo circle or part of it (an arc), which is quite possible close to the full moon (which was on Dec 16).
Furthermore, {\em white qi} is reported among some stars of Taurus, which could well be some
other part of the lunar halo display (e.g. an arc of the halo ring), as it is reported to have
{\em followed the moon}. Because all these objects  (moon, Pleiades, and Taurus) were located in
the east and south in the first half of the night, and because the reported {\em qi} was always
close to the (almost full) moon, it cannot have been an aurora.
This event is not listed as an aurora in Yau et al. (1995) nor in Xu et al. (2000).

Keimatsu (1973) translated as follows:
\begin{quotation} 
... the moon covered the Mao ... the Pi [in Tau] put on a white halo -- and a mass
of white vapour coming from the north penetrated the Mao
\end{quotation}
and he then classified this event as a {\em probable} aurora.
Keimatsu understands the object of the coverb {\em cong} as {\em north} and renders it as a preposition, 
hence {\em from the north}. We understand {\em cong} as a full verb meaning {\em follow} here, 
with the implied object being the moon. Keimatsu's rendering works grammatically, 
but given that the moon is already said to be surrounded by a white halo, 
we think it is more likely that the {\em qi} is moving with the moon rather than coming from the north.

The event is also listed as as aurora in Silverman (1998, 
see also online catalogue\footnote{nssdcftp.gsfc.nasa.gov/miscellaneous/aurora, 
note that this catalog lists {\em sources} of aurora reports instead of aurora {\em events},
so that events found in several sources are listed several times.}).

\subsection{AD 763 Nov: {\em purple qi ... met their horses' heads}}

For AD 763 Nov, we find in Su E's (fl. AD 886) {\em Duyang zabian} (Duyang Miscellany) (1.1) for China: 
\begin{quotation}
In the first year of the Guangde reign period 
of the Daizong Emperor [AD 763], Tibetan forces invaded the Bian Gate Bridge.  
The Emperor visited Shan County. The royal armies found themselves at a disadvantage, 
as there was constantly a purple qi like the canopies of chariots that met their horses' heads.  
Once they returned to Tong Pass, the Emperor sighed and said: 
How broad and vast are the waters that send me off to the east ...
\end{quotation}

There is no direct indication that the purple {\em qi} here was seen in the sky, 
or even at night or in the north, much less that it was an aurora. 
On the contrary, the text specifies that the {\em qi} was 
constant and on the same level as the horses.
It was most likely fog.
This event was misinterpreted as an aurora in Dai \& Chen (1980) 
and also listed as an aurora in Silverman's online catalog for AD 763 Nov (for Tongguan, China).
The event is neither listed in Keimatsu (1973) nor Yau et al. (1995) nor in Xu et al. (2000).

\subsection{AD 767 Aug 25: {\em white qi permeating the sky}}

For AD 767 Aug 25, we find in {\em Jiu Tang shu} 11.287 for Shaanxi, China:
\begin{quotation}
[In the 7th month of the 2nd year of the Dali reign period,] on 
the jiaxu day [AD 767 Aug 25] during the you hour [17-19h], 
there was white qi permeating the sky.
\end{quotation}

Keimatsu (1973) translated as {\em ... white vapour spreading over the sky} 
and then classified it as {\em probable to doubtful}. 
At the capital city of Xi'an in Shaanxi, China,
sunset on AD 767 Aug 25 (on the Julian calendar) was at about 18:30h, 
end of civil twilight at about 18:56h, 
and the end of astronomical twilight was even at about 19:58h (local times).
An aurora observation is extremly unlikely for civil twilight and the first few minutes 
of nautical twilight. Hence, this is most certainly not an aurora.
This event is not listed in Yau et al. (1995) nor in Xu et al. (2000),
but is included in Usoskin et al. (2013).\footnote{Usoskin et al. (2013) wrote after citing
Keimatsu (1973) and Yau et al. (1995):
{\em The next nearest observations are at AD 767 and 786.}, 
so that they included one or more Chinese events in AD 767,
probably all three;
we discuss all three such events here, Sect. 2.3, 2.4, and 4.3.}

\subsection{AD 767 Jul/Aug: {\em three stars drop down}}

For AD 767 (between Jul 30 and Aug 28, most likely Aug 25), we find in the {\em Silla pon'gi} 
(Annals of the Kingdom of Silla, trad. 58 BC-AD 935) section of the {\em Samguk sagi} 
(Historical Records of the Three Dynasties) (9.95), compiled by Kim Pusik (AD 1075-1151) and others during the reign 
of Injong (AD 1122-1146), the following event in Korea: 
\begin{quotation}
In the autumnal seventh month of the third year of King Hyegong (r. 765-780)
... while the King was attending a banquet in the Hall of State Ceremonies, 
there appeared three stars which fell and collided against one another above the Royal Court.  
Their light was like fire as it advanced and dispersed.
\end{quotation}

This is probably listed as an aurora in Usoskin et al. (2013), see footnote no. 4, 
though it is classified as {\em doubtful} in Keimatsu (1973),
who translated {\em ... three stars drop down ...}.
The three stars which fell were almost certainly meteors, most likely the Perseids.
This event is not listed in Yau et al. (1995) nor in Xu et al. (2000).

\subsection{AD 773 Aug 9: {\em 17-19h ... bands of qi}}

For AD 773 Aug 9, we find in {\it Jiu Tang shu} 36.1327 (Liu 1975) for Shaanxi, China (full moon on Aug 6/7): 
\begin{quotation}
Between 17-19h ({\em you} double-hour), there were three bands of qi strung across the sky.
\end{quotation}

Keimatsu (1973) interpreted this report as a {\em probable to doubtful} aurora; 
Usoskin et al. (2013) also listed this event ({\em credible observations from Shanxi Province, China, 
in AD 770 (twice), AD 773, and AD 775. The next nearest observations are at AD 767 and AD 786}). 
At the city of Xi'an in Shaanxi,\footnote{The transliteration by Keimatsu (1973)
and Usoskin et al. (2013), namely {\em Shanxi}, is misleading, because the 
old Chinese capital Xi'an is located in the {\em Shaanxi Province}
next to another province called {\em Shanxi Province}.} China, sunset on AD 773 Aug 9 
was at about 18:49h and the end of civil twilight was at about 19:16h (local times).
Hence, it was observed before the end of civil twilight, probably even before sunset;
therefore, the event was not an aurora. It may have been a solar halo display
with three pillars of light as {\em three bands of qi strung across the sky}.
This report is not listed in the aurora catalog of Yau et al. (1995) nor in Xu et al. (2000).

\subsection{AD 774 Oct 13: {\em Red Drops from the ... northeast}}

In the astronomical treatise of {\em Jiu Tang shu} 36.1328, 
an event is listed for AD 774 Oct 13 for China as follows:
\begin{quotation}
[(a)] In the ninth month [of the ninth year of the Dali reign period] on the gengzi day [AD 774 Oct 13],
the 36th day of the sexagenary cycle, Red Drops from the lands in the northeast (Youzhou) entered the
morning light (chao). That night Venus entered the Nandou asterism [Sgr].
\end{quotation}

There is an additional instance of {\em Red Drops} in 36.1327 in connection with the northeast for AD 774: 
\begin{quotation}
[(b)] In the night of the wuwu day of the eighth month [no wuwu day in the eighth month, AD 774], 
Mars approached the moon. In that month, Red Drops came from the Youzhou [in the northeast] to enter the court.
\end{quotation}

As the events are recorded in an astronomical treatise, one might suspect that {\em Red Drops} refer to meteors or aurorae.
However, Red Drops (AD 742-784) in fact turns out to be the unusual sounding name of a Northeastern frontier general
who rebelled soon after visiting the imperial court (or {\em chao}), which is referred to by the same term
as {\em morning light} due to the fact that it is held in the morning. Hence, this was a political event
rather than an astronomical phenomenon -- but as it was listed in the astronomical treatise in the {\em Jiu Tang shu}
(probably because the Moon, Mars, Venus, and a constellation are mentioned in association with it).
We include it here for completeness and to avoid a misinterpretation in the future.

\subsection{AD 775 Dec 31 or AD 776 Jan 12: {\em white qi above the moon in the east} ?}

This event has traditionally been given two different dates, 
having occurred on a {\it jiazi day}, the first day in the sexagenary cycle, 
according to premodern editions of the {\it Jiu Tang shu} and on a {\it bingzi} day, 
the 13th day of the sexagenary cycle, according to the {\it Xin Tang shu}.

In premodern editions of the {\it Jiu Tang shu}, we can find the following text for Xi'an, China:
\begin{quotation}
On the night of the jiazi day of the twelfth month, 
above the moon in the east, there were more than ten bands of white qi like a bolt of silk, 
penetrating Wuche (Aur), Dongjing (Gem), Yugui (Cnc), Zui (Ori), Shen (Ori), Bi (Tau), 
Liu (Hya), and Xuanyuan (Lyn, Leo, and LMi). 
Just after the third watch, they vanished.
({\em Jiu Tang shu} 36.22/Liu 2007)
\end{quotation}

The modern standard edition of the {\it Jiu Tang shu}, Liu 1975 (36.1328), 
gives the date as {\em bingzi}, so that it accords with Ouyang Xiu's (AD 1007-1072) 
{\it Xin Tang shu} (New History of the Tang Dynasty; 1975), 
Wang Pu's (AD 922-982) {\it Tang hui yao} (Essential Records of the Tang Dynasty; 2007), 
and Ma Duanlin's (ca. AD 1254-ca. 1323) {\it Wenxian tongkao}
(Comprehensive Investigation of Historical Documents; 1986). 

The later {\it Xin Tang shu} (32.836/Ouyang 1975) uses a different phraseology: 
\begin{quotation}
On the bingzi day of the twelfth month, the moon rose in the east, 
and above it there were more than ten bands of white qi, 
like bolts of bleached silk, penetrating Wuche and Bi, Zui, Shen, Dongjing, 
Yugui, Liu, and Xuanyuan. In the middle of the night it dispersed.
\end{quotation}

Keimatsu (1973) listed this report as a {\em very probable aurora}.
While he gave two text variants with two different dates in Chinese script,
he dated the event to AD 775 Dec 31 (a {\it jiazi} day, i.e. 1st day of the sexagenary cycle) 
in his English translation
based on the premodern text of the {\it Jiu Tang shu}.
Neither Yau et al. (1995) nor Xu et al. (2000) list it as an aurora.
However, since the new moon was on AD 775 Dec 26/27, the 4-day-old moon could not be seen 
in the east on AD 775 Dec 31.
This discrepancy can be solved by using the date given in
the modern standard {\em Zhonghua} edition, AD 776 Jan 12 (a {\em bingzi} day),
which is consistent with the lunar phase: 
The full moon was on AD 776 Jan 10/11, 
so that the moon was in the east in the first half of the night 
(moonrise at Xi'an on AD 776 Jan 12 was at around 19:40h local time).

All constellations mentioned rise in the first half of the night from the east and 
move towards the south with the moon. 
The reported phenomenon is located only and exactly {\em above the moon}
and nowhere else
-- apparently even comoving with the moon and the stars;
the moon was located in (and around) Leo on Jan 12.
This scenario is not consistent with aurorae
(partly because it is too bright close to the moon).
The {\em bands of white qi, like bolts of bleached silk} may have been a lunar halo effect.
 
Considering all events listed in the catalogues of Yau et al. (1995), Xu et al. (2000), 
and Keimatsu (1973, 1974) for several centuries (AD 550 - 1006), 
there is only one case where the {\em moon}
was mentioned {\em expressis verbis} in the context of likely true aurora 
(AD 1006 Apr 14, Xu et al. 1995): {\em In the north, a scarlet vapour extended
across the sky, and a white vapour penetrated the moon}, where the first (scarlet)
vapour is a reliable aurora in the north, while the {\em white vapour} may be
a halo effect near full moon (Apr 16).

The Jan 776 event 
is also misdated in the aurora catalog of Silverman (1998) following Keimatsu (1973).

Zhou et al. (2014) listed this particular event as as aurora for 
{\em 11 Dec 774, i.e. 17 January AD 775} 
with a slightly different English translation. 
Apart from the fact that we do not understand the two dates 
(both of which err by about one year), 
the duration and the hours of the sighting are not correct:
The nearly full moon on the true date of the sighting (AD 776 Jan 12) 
was visible at the relevant location from around 19:40h local time.
As the record 
specifies that the 
{\em ten band of white qi}
were visible 
{\em above the moon}
in the east, 
it is unlikely that the sighting could have started as early as 17-18h,
as claimed by Zhou et al. (2014). 
The records are somewhat vague regarding the time when the phenomenon could no longer be seen.  
The {\it Jiu Tang shu} does specify that it vanished {\em just after the third watch}.
For Xi'an, China, on the night of AD 776 Jan 12/13, the third night-watch ran 
from about 22:53h to 01:25h (local time),\footnote{Given the necessarily speculative 
discussion in Sect. 5.2 regarding a connection
with an event on Jan 11 (an {\em yihai} day, the 12th day of the sexagenary cycle), 
it could be that {\em bingzi} here means
the {\em portion of the night after midnight} 
and {\em yihai} meaning the {\em portion of the night before midnight};
hence, the event could conceivably have occurred 
on the night of AD 776 Jan 11/12;
considering the {\em bingzi} date, it could be the night of AD 776 Jan 12/13.
}
{\em just after the third watch} suggests some time just after 1:25h;
this is not inconsistent with the {\it Xin Tang shu}, which states 
that it dispersed in the {\em middle of the night}.  
Given these somewhat ill-defined parameters, 
the phenomenon was most likely visible for less than $\sim 6$h.  
The records do not support any greater precision.   

The massive late imperial encyclopaedia {\em Gujin tushu jicheng}
(Collectanea of Ancient and Modern Charts and Books; compiled 1726-8) 
contains a report which conflates the appearance of white {\em qi} 
above the moon
in AD 776 Jan with an appearance of a comet (see Sect. 5.2).

Stephenson's (2015) translation of the Jan 776 entry agrees with our own. 
He first comments that he has {\em little doubt that an auroral display is described here} 
and classifies it as the only {\em confidently identified ... auroral sighting} 
in his section 5.1.   In his section 5.3, he describes it
as the {\em single definite record of an auroral display}.  
In his section 6, he calls it the {\em single reliable East Asian report 
of the aurora}. Cliver (2014) reported: {\em From a re-examination
of Keimatsu (1973), F.R. Stephenson (2013, priv. comm.)
... suggests ... that a report of more than 10 bands of white
vapour on AD 776 January 12 may be auroral in nature} (Cliver et al. 2014). 
Serious doubts persist, however, concerning the identification of this event as an aurora.  
Stephenson (2015) did not discuss why it was not classified as aurora in Yau, Stephenson
\& Willis (1995), nor did he consider the halo hypothesis,
or the fact that the white phenomenon appeared in
the eastern to southern direction, not in the north.

Regarding the question of whether this event was observed on the {\em jiazi}
or {\em bingzi} day, Stephenson (2015) writes that Keimatsu
{\em mistakenly cites the day ... jiazi rather than bingzi},
but Keimatsu in fact gave both text variants in Chinese script
(also shown here at the bottom our Fig. 1);
Stephenson (2015) did not discuss the inconsistent lunar phase on one of the two 
dates, nor did he specify any other reason as to why the {\em jiazi} day was wrong.

\section{Likely true Chinese aurorae observations from AD 757 to the 770s}

Here, we briefly cite the Chinese reports of likely true aurorae in the AD 760s and 770s,
in order to show how reports of true aurorae read. Our translations are provided below,
the Chinese texts in Fig. 2.

These reports mention night-time, northern direction, as well as red colour,
all typical for very likely true aurorae.
Most reports are dated within a few days around the new moon, when it is very dark at night.
Three examples of very likely true aurorae occurred in AD 762.

\begin{figure*}
{\includegraphics[angle=0,width=17cm]{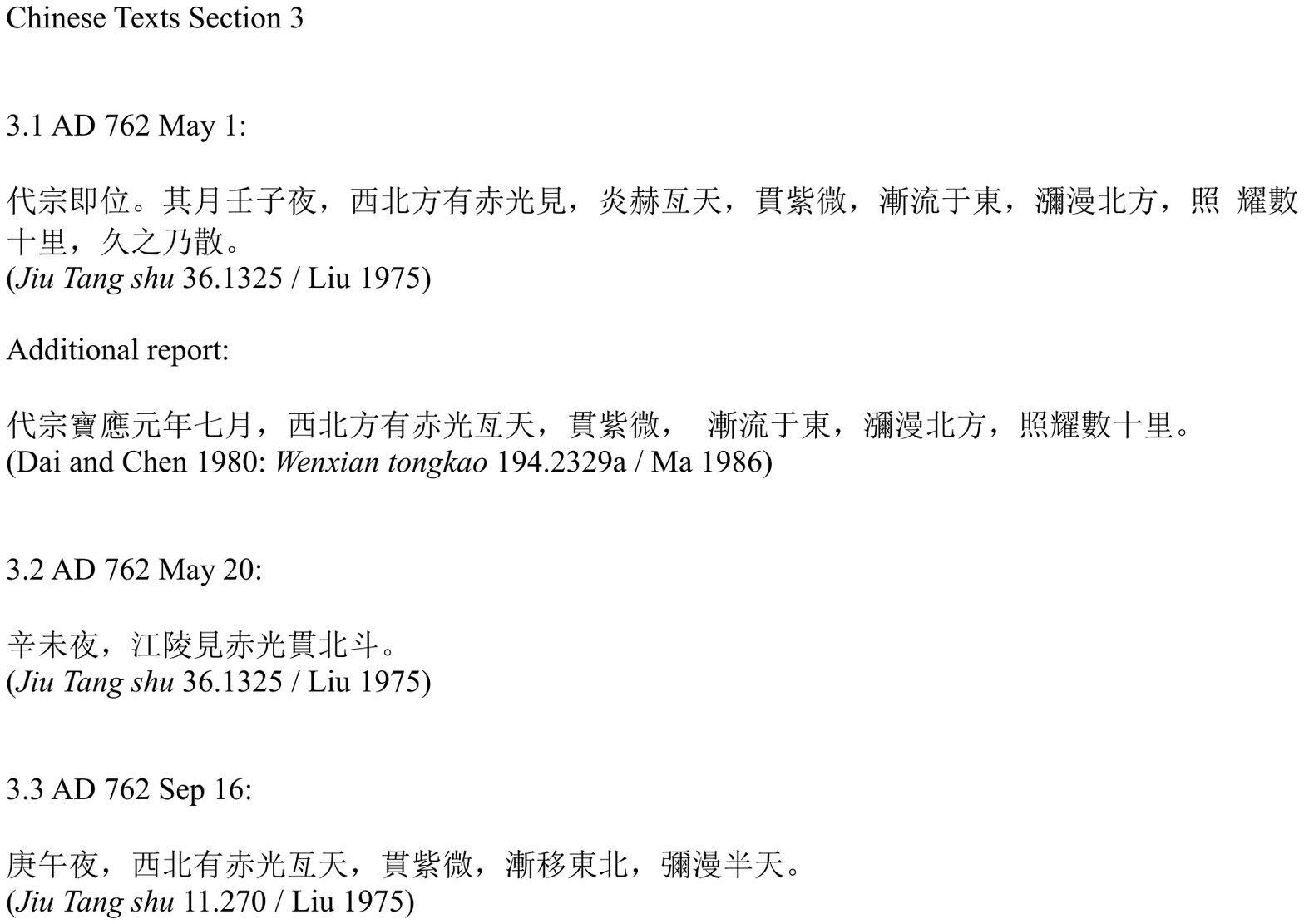}}
\caption{Here, we show the Chinese texts related to likely true aurorae. 
Translation and discussion are given in the Sect. 3.}
\end{figure*}

\subsection{AD 762 May 1: {\em on the night ... red light ... NW}}

AD 762 May 1, Shaanxi, China (new moon Apr 28/29): 
\begin{quotation}
The Daizong Emperor took the throne. In the same month, on the night of renzi 
[in the first year of the Baoying reign period, AD 762 May 1], 
a red light appeared in the northwest. 
Its flames shined bright and stretched across the sky, penetrating Ziwei [the circumpolar region].
It gradually flo\-wed eastward, slowly filling the north. It shined round for several tens of li, 
and after a long time, it dispersed.
\end{quotation}
From {\em Jiu Tang shu} 36.1325, also as aurora in Keimatsu (1973)
({\em certain} aurora), Yau et al. (1995), and Xu et al. (2000). 

Yau et al. (1995) gave the date as the {\em second year} of the Baoying reign period (AD 762). 
The discrepancy in our translations arises from an anomalous situation during 
the transition from the reign of Emperor Suzong (r. AD 756-761) to the reign
of Emperor Daizong (r. AD 762-779).
The previous reign period, Shangyuan, had been ended but not been replaced 
in the ninth month of its second year, AD 761 Oct 3-Nov 1.  
The Baoying period did not in fact begin until the fourth month of the next year, AD 762 Apr 29-May 27.  
Thus, there is a gap between the two reign periods.  
Yau et al. (1995) seem to treat the end of the Shangyuan reign period as 
if it were the beginning of the Baoying, and thus treat AD 762 as its second year.
(This also holds for the next events in AD 762, Sect. 3.2 and 3.3.) 

As far as the {\em li} is concerned, the actual distance varied.  
It was equivalent to 300 {\em paces} or 1800 Chinese feet in Tang times, roughly one-third of a mile.
Hence, the total estimated length was several times 5-6 km (one Chinese foot being 30.3 cm). 
(To estimate the linear length, one would of course need to know the distance,
which was probably very roughly estimated.) 
The large size given suggests an easily visible, impressive phenomenon. 
Yau et al. (1995) translated the word {\em he} as {\em blazing flames};
it may indicate the color red, an appearance of anger, something that is eye-catching, 
or act as a verb meaning {\em shine brightly}; we prefer {\em its flames shined bright}. 

There is another report with almost the same wording -- hence, probably about the same event 
(but leaving out the exact date and that it occurred at night) at a slightly later time in the same year,
AD 762 Jul/Aug (between Jul 26 and Aug 23), China: 
\begin{quotation}
In the seventh month of the 1st year of the Baoying reign period of the Daizong Emperor [AD 762 Jul 26 to Aug 23],  
there was a red light in the northwest that stretched across the sky, penetrating Ziwei [the circumpolar region].  
It gradually flowed eastward, slowly filling the north. It shined round for several tens of li.
\end{quotation}
From {\em Wenxian tongkao} (AD 1254-1323), also listed in Dai \& Chen (1980) and the Silverman online catalog
(for AD 762 Aug for Xi'an, China).  

The language of the passage largely matches that of the AD 762 May 1 event as recorded in {\em Jiu Tang shu} 36.1325.
The wording also overlaps, partially, with the AD 762 Sep 16 event.
The {\em Wenxian tongkao} passage (also cited in Dai \& Chen 1980) uses language almost identical to 
that of the {\em Jiu Tang shu},
but places the event in the seventh month of the same year (AD 762 Jul 26 to Aug 23).  
However, this date is perhaps due to copyist error of two nearly homophonous characters.  
Whereas the {\em Jiu Tang shu} has the phrase {\em qi yue} meaning {\em the month of his accession}. the 
{\em Wenxian tongkao} has {\em qi yue} meaning {\em the seventh month}. 
The two {\em qi} characters 
were pronounced in different tones, however, so our hypothesis remains tentative.
In any case, there were credible aurorae in the year AD 762.

\subsection{AD 762 May 20: {\em On the night ... red light}}

AD 762 May 20, Jiangling county, Hubei, China (new moon May 27/28): 
\begin{quotation}
On the night of xinwei [in the fourth month of the 1st year of the Baoying reign period, AD 762 May 20], 
in Jiangling a red light appeared which penetrated Beidou [the Big Dipper in UMa, i.e. north].
\end{quotation}
From {\em Jiu Tang shu} 36.1325, also listed as an aurora in Kei\-ma\-tsu (1973)
({\em very probable}), Yau et al. (1995), and Xu et al. (2000);
Yau et al. (1995) again had the 2nd year of the Baoying reign period, AD 762, see comment in Sect. 3.2.

\subsection{AD 762 Sep 16: {\em On the night ... NW ... red light}}

AD 762 Sep 16, Shaanxi, China (new moon on Sep 22):
\begin{quotation}
On the night of gengwu [in the eight month or the 1st year of the Baoying reign period, AD 762 Sep 16], 
in the northwest there was a red light that stretched across the sky, penetrating Ziwei [circumpolar]. 
It gradually moved northeast, until it filled half the sky.
\end{quotation}
From {\em Jiu Tang shu} 11.270, also listed as a certain aurora in Keimatsu (1973),
Yau et al. (1995), and Xu et al. (2000).
In the Yau et al. (1995) and Xu et al. (2000) translation, only the NE direction is mentioned, 
not the NW, but both NW and NE directions are clearly in the original Chinese text;
Yau et al. (1995) again had the 2nd year of the Baoying reign period, AD 762, see comment in Sect. 3.2. 

This text from Xi'an, the modern capital of Shaanxi, written by official court astronomers, shows how systematic
the night reports are, listing the date, the fact that it occurred at night, its direction in the sky, 
colour, area in the sky, motion, and extent -- similar to the {\em Jiu Tang shu} report in Sect. 3.1.
The text partly overlaps with the {\em Wenxian tongkao} report given in Sect. 3.1, and may be a duplication.

\section{Questionable aurora reports AD 757-779}

Now, we also list those celestial observations from East Asia from AD 757 to the end of the 770s,
where the interpretation is uncertain, so that they could either be aurora or something else.
Again, we give our translations, and the Chinese texts can be found in Fig. 3.

\begin{figure*}
{\includegraphics[angle=0,width=17cm]{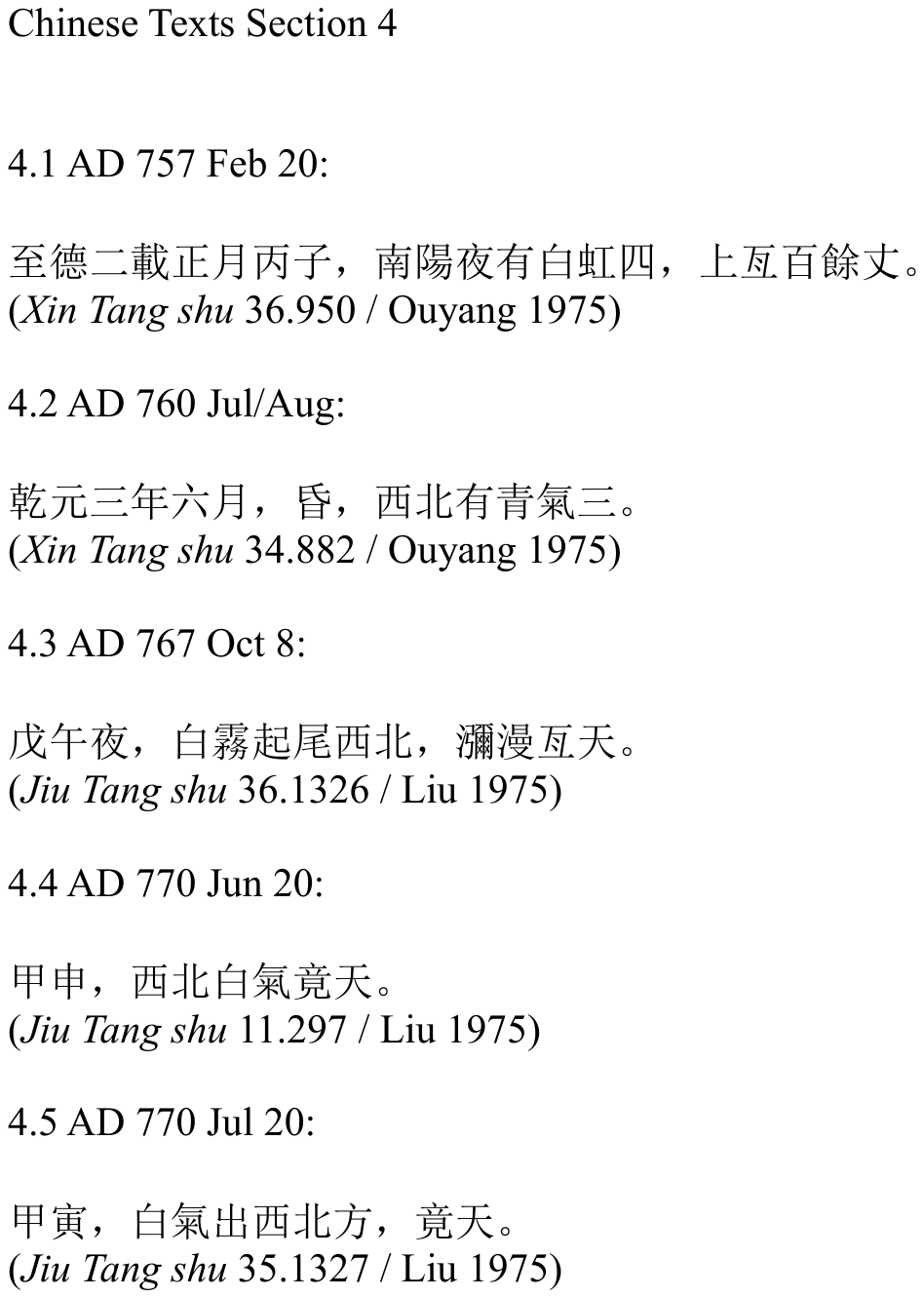}}
\caption{Here, we show the Chinese texts related to questionable cases, which may or may not be aurorae.
Translation and discussion are given in the Sect. 4.}
\end{figure*}

\subsection{AD 757 Feb 20: {\em in the night ... four white rainbows}}

AD 757 Feb 20, Nanyang, China (new moon Feb 22/23): 
\begin{quotation}
On the bingzi day [13] of the first month of the second year of the Zhide reign period [AD 757 Feb 20],
in the night, at Nanyang, there were four white rainbows [hong].
These extended upwards for more than one hundred zhang.
\end{quotation}
From {\em Xin Tang shu} 36.950, 
also listed as an aurora in Kei\-ma\-tsu (1973) ({\em probable to doubtful}), 
Yau et al. (1995), and Xu et al. (2000). 

There are several different words conventionally translated as {\em rainbow} in classical Chinese.
{\em Daidong} appears in the {\em Odes classic} ({\em Shijing},
while both {\em ni} and {\em hong} appear in astronomic and omenological texts,
where they are largely used synonymously.
A distinction is sometimes drawn in the genders of the {\em ni} and the {\em hong};
the term {\em ni} often occurs in the phrase {\em ci ni}, a female rainbow.
These words should perhaps be construed somewhat more broadly
than the English {\em rainbow}, as they can occur at night,
and can be white rather than multi-coloured.
The entry in {\em Xin Tang shu} 36.950 for the AD 757 Feb 20 event 
uses the single most common of these, {\em hong}.
Unlike the English word {\em rainbow}, {\em hong} is composed of a single element, 
not two independently meaningful morphemes, {\em rain} and {\em bow}.
There is no commonly used single word meaning {\em night rainbow} in classical Chinese.
One might suspect that the word for {\em bow} would be used for {\em halos}, 
but the word for {\em bow}, as in a bow and arrow, is {\em gong}.
It is quite distinct from {\em hong}, both in sound and graphic form,
and does not normally enter into astronomical discourse.  
The most common word for {\em halo} is perhaps {\em huan}, which in common discourse simply means {\em ring}.  

While the term {\em hong} usually refers to a circular or semi-circular glow,
the event cannot be a lunar halo display nor a night rainbow, because it was near the new moon,
and it could not have been solar halo display, because it was at night.
A {\em zhang} is ten Chinese feet, i.e. 303 cm in Tang times 
for general measurements, or ten degrees in astronomy (Wilkinson 2000);
since a celestial phenomenon cannot have the size of 1000 degrees ({\em one hundred zhang}) 
or more, the word {\em zhang} here stands for the linear length (ten Chinese feet).
The wording {\em more than 100 zhang} refers to the height -- being more than 300 m.
The report is not from the capital and is listed in a treatise on general 
omenology rather than astronomy.
The text as it stands does not allow us to identify the precise nature of the phenomenon 
and caution is warranted in its interpretation.

\subsection{AD 760 Jul/Aug: {\em northwest ... three green qi at dusk}}

AD 760, between 17 July and 15 August, China:
\begin{quotation}
In the sixth month of the third year of the Qianyuan reign period,
in the northwest there were three green qi at dusk.
\end{quotation}
From {\em Xin Tang shu} 34.882, also as aurora in
Dai \& Chen (1980), Yau et al. (1995), and Xu et al. (2000). 

{\em Dusk} is more or less equivalent to the Chinese word it renders, {\em hun}.
It implies that it is already getting dark and generally indicates the time between sunset and nightfall.
If the {\em hun} period means those 36 min mentioned in the introduction between sunset 
and the start of the first night-watch (Stevenson 1997), then the event would have occurred 
(at least mostly) during civil twilight, when it is still too bright for an aurora. 

{\em Qi} can refer to clouds or aurorae. This record occurs in a treatise on omenology.
The colour {\em qing} could refer to black, blue, or green colour.
Given that the {\em qi} occurred in the northwest and had colour, 
it is possible that this was a blue or green aurora;
it could have been dark enough for an aurora;
it could also be some kind of a halo effect at or shortly after sunset (NW).

\subsection{AD 767 Oct 8: {\em On the night ... white mist ... NW}}

AD 767 Oct 8, Shaanxi, China (full moon Oct 12): 
\begin{quotation}
On the night of wuwu [AD 767 Oct 8], white mist rose up to Wei [in Scorpio] 
in the northwest, spreading across the sky.
\end{quotation}
Keimatsu (1973) classified the event as a {\em doubtful to probable} aurora;
the event is also listed as aurora in Yau et al. (1995) and -- probably -- also
in Usoskin et al. (2013), but not in Xu et al. (2000). 

The word for mist usually means fog ({\em wu}), so that the classification as an aurora is uncertain.
The phenomenon was seen projected onto a stellar constellation at night.
However, as we can see in the chapters in Qutan Xida's (fl. 729) {\em Kaiyuan zhan jing} (Kaiyuan reign period [713-741] Classic of Prognostication) 
on omens of clouds and {\em qi} encroaching on constellations, odd clouds or other meteorological phenomena 
can appear in particular constellations.
The title of chapter 95 reads, for instance, {\em Yunqi fan ershiba xiu zhan} or {\em Prognostications on Clouds and Qi Transgressing the 28 Lunar Lodges},
the 28 lunar lodges being the constellations that make up the zodiacal band in Chinese astronomy.

An excavated manuscript from the Library Cave (Mogao ku) at Dunhuang (S. 3326), 
often referred to as the world's earliest extant star atlas, 
likewise suggests that clouds or misty formations comprised of various colours of {\em qi} were read against constellations 
for the purposes of prognostication. The manuscript, produced between AD 649 and AD 684, contains a series of images of variously 
shaped cloud-like formations, a chart depicting some 257 asterisms containing 1,339 stars, and a robed figure brandishing a 
bow and arrow labeled ``God of Lightning'' (Bonnet-Bidaud, Praderie, and Whitfield 2009).  
Scholarly interpretation of the manuscript has focused on its astronomical features and little attention has been paid to 
the portion of the manuscript containing the cloud-like formations. Both Bonnet-Bidaud, Praderie, and Whitfield (2009) 
and Sun \& Kistemaker (1997) present images of the portion of the manuscript containing the star chart, 
but neither present images of the cloud-like formations.  
However, when examined as a whole, the manuscript quite clearly shows the extent to which 
astronomical and meteorological phenomena were mutually integrated in medieval China.

\subsection{AD 770 Jun 20: {\em NW ... white qi ... across the sky}}

AD 770 Jun 20, Shaanxi, China (moon's last quarter Jun 20/21): 
\begin{quotation}
On the jiashen [21] day [of the fifth month of the fifth year of the Dali reign period],
in the northwestern direction, white qi permeated the sky.
\end{quotation}
From {\em Jiu Tang shu} 11.297, also in Yau et al. (1995) and Keimatsu (1973)
({\em probable to doubtful}),
but not in Xu et al. (2000), because they omit all events, where it is not
mentioned explicitly that they were observed at night-time.
See the discussion in the next subsection.

\subsection{AD 770 Jul 20: {\em NW ... white qi ... across the sky}}

AD 770, Jul 20, Shaanxi, China (moon's last quarter Jul 20): 
\begin{quotation}
On the jiayin [51] day [of the sixth month of the fifth year of the Dali reign period], 
white qi emerged in the northwest and permeated the sky.
\end{quotation}
From {\em Jiu Tang shu} 35.1327, also in Yau et al. (1995) and Keimatsu (1973) ({\em probable to doubtful}),
but not in Xu et al. (2000).  

It is likely that both reports refer to the very same event and that one of the two
reports is a misdated copy of the other. The dates differ by exactly one month.
The day in the sexagenary system is given as {\em jia-shen} (21) in the first report (i.e. AD 770 Jun 20)
and {\em jia-yin} (51) in the 2nd report (i.e. AD 770 Jul 20); both words consist of two parts,
the first parts being identical. The graph for {\em shen} is quite similar to the lower portion 
of the graph for {\em yin}, and the two words could easily be mistaken for one another in a manuscript.  
However, the final redactions of the two passages seem to have taken this into account, 
appropriately placing the two entries in the 5th and 6th months.
The former text is from the Basic Annals ({\em Ben ji}) of the reign of Emperor Daizong 
in the {\em Jiu Tang shu} (11.297), while the latter is from a treatise on general omenology, 
rather than celestial patterns, in the same larger historiographical project (35.1327).  
The discrepancy between the two passages might be explained in a number of ways: 
One passage might be an inaccurate paraphrase of the other; 
the two passages might represent varying reproductions of the same damaged manuscript; 
or the two passages might be based on two different manuscripts representing two different lines 
of transmission of the same original record.  
Neither is it beyond the realm of possibility that the two passages were in fact 
records of two independent events. Assuming they are records of the same event,
it is impossible to determine which date is correct, if one of the two reports is a misdated copy of the other. 

Since the reports for the last two events (AD 770 June and July) do not specify night-time,
it is also not certain that they refer to aurorae.
However, given the (slightly) different texts for June and July 770,
we consider it possible that they refer to two distinct but similar events, possibly aurorae
({\rm northwest} direction and {\em qi ... across the sky}).
Given that about one month was between these two events, they could refer to auroral activity
from a stable coronal hole. 

\section{Chinese comet observations reported for the AD 770s}

Since a comet observation was also discussed in the context of the $^{14}$C 
variation around AD 775 (Liu et al. 2014), 
we also briefly review Chinese comet observations from the mid AD 770s. 

\begin{figure*}
{\includegraphics[angle=0,width=17cm]{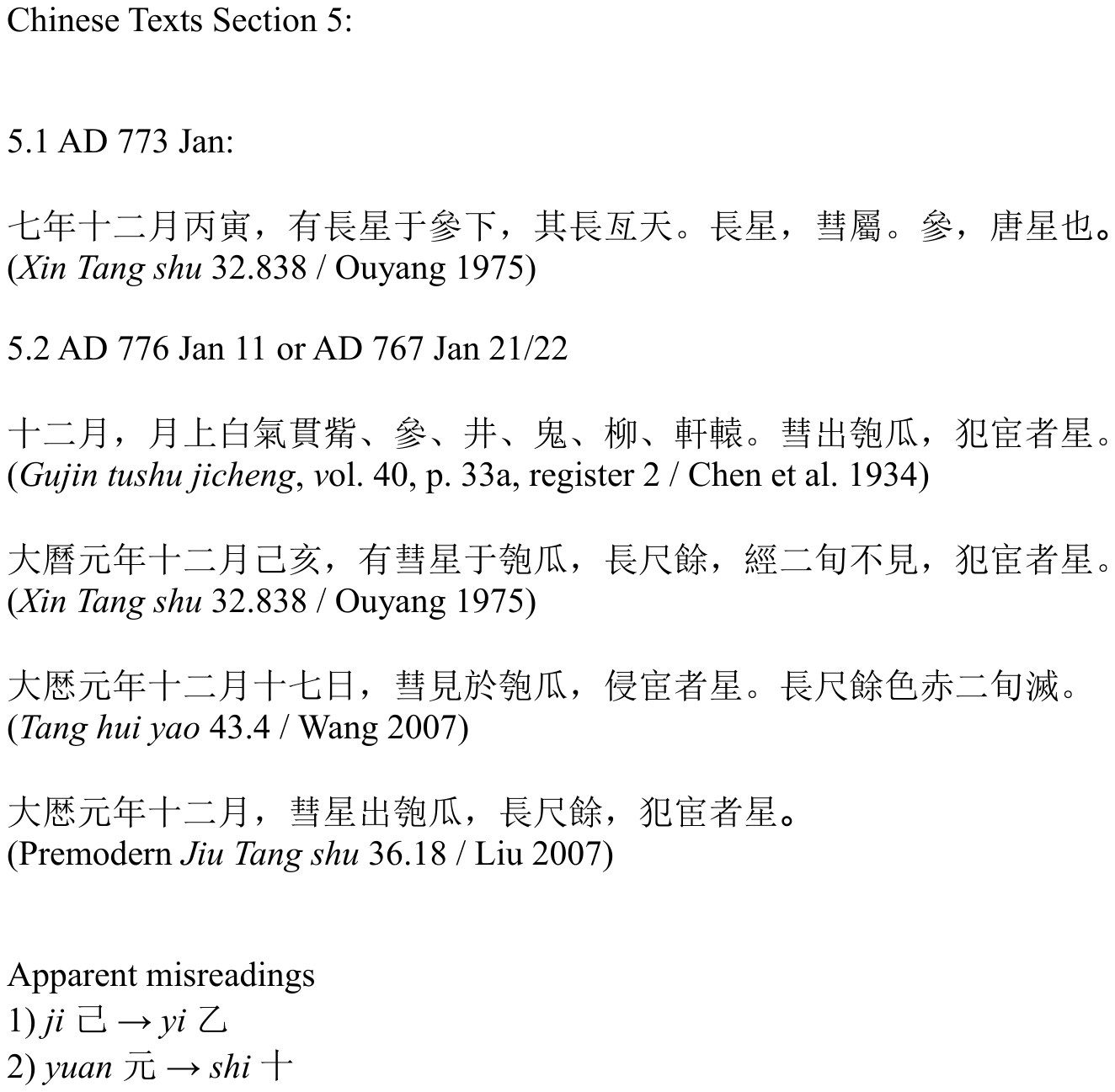}}
\caption{Here, we show the Chinese texts related to comets.
Translation and discussion are given in the Sect. 5.
The apparent misreadings at the bottom are discussed in Sect. 5.2.}
\end{figure*}

\subsection{The comet of AD 773 January}

Liu et al. (2014) claimed that the Chinese had observed 
a comet that had collided with the Earth's atmosphere 
on AD 773 Jan 17, thus bringing extra $^{14}$C into the atmosphere. 

However, as we have shown already (Chapman et al. 2014), 
this claim was not correct for several reasons: 
\begin{enumerate}
\item The Chinese thought at that time that comets were inside the Earth's atmosphere, 
so that they could not interpret a comet observation as a collision with Earth.
\item The Chinese report presented in Liu et al. (2014) 
was culled from several different sources and inaccurately translated. 
The correct translation of the most extensive version of the record is: \\
{\em On the bingyin day of the twelfth month of the seventh year (AD 773 Jan 17), 
there was a long star beneath Shen. Its length extended across the sky. 
Long stars belong to the class of comets. Shen is the constellation corresponding to the Tang}
({\it Xin Tang shu} 32.838).
\item The comet was then observed by the Japanese on AD 773 Jan 20, 
so that it could not have collided with the Earth on Jan 17.
\end{enumerate}
See Chapman et al. (2014) for details.
This case was re-examined by Stephenson (2015)
and fully confirmed.

\subsection{The presumable comet of AD 776}

Next, we discuss the presumable comet of AD 776,
partly because it is reported in connection with the white {\em qi} phenomenon,
which may be the same as the one identified by Stephenson as an aurora.

It is highly probable that the comet Hasegawa (1980) cites as appearing 
on AD 776 Jan 11 (no. 437) is spurious, 
the result of a mistranscribed date.  
The details concerning the comet correspond 
very closely with those of an earlier comet appearing on AD 767 Jan 21 (no. 433).

Hasegawa (1980) cites the following passage in the  
18th century encyclopedia {\it Gujin tushu jicheng}
as the source for comet  no. 437,
{\em Gujin tushu jicheng}, vol. 40, p. 33a, register 2 (Chen et al. 1934): 
\begin{quotation}
In the 12th month [of the 10th year of the Dali reign period, AD 776 Jan], 
white qi above the moon penetrated Zui, Shen, Jing (Gem), Gui (Cnc), Liu, and Xuanyuan.  
A comet appeared in Hu Gua (Del). It fell into Huanzhe (Her).
\end{quotation}

In his notes on the comet no. 437, Hasegawa (1980) cites an unspecified Chinese record: 
\begin{quotation}
A Chinese record says: On a yihai day in the 12th month of the tenth year 
of the Dali reign-period [AD 776 Jan 11], a comet appeared at Hu Gua ($\alpha$, $\beta$, $\nu$ Del). 
It was several feet in length, and after 20 days it disappeared. 
It fell into Huanzhe (near $\alpha$ Her).
\end{quotation}

The {\em Gujin tushu jicheng} entry seems to conflate two separate events: 
An appearance of white {\em qi} above the moon 
in AD 776 Jan 
(c.f. Sect. 2.7), 
and the appearance of a comet (Hasegawa comet no. 433),
whose correct date is AD 767 Jan 21/22, as we will show below.
Although the {\em Gujin tushu jicheng} entry does not specify a sexagenary date, the two events
may have been conflated due to confusion between the {\em yihai} and {\em jihai} dates in the sexagenary calendar 
coupled with mistranscription of the year, as both occurred in 12th month of years in the Dali reign period.  
The {\em yihai} day 
(from Hasegawa's notes, see above)
is the 12th day in the sexagenary calendar 
and immediately precedes the {\em bingzi} day (on which the appearance 
of white {\em qi} in Sect. 2.7 was said to have occurred.) {\em Jihai} 
in contrast is the 36th day in the sexagenary calendar and is given 
as the date for the appearance of the AD 767 Jan 22 comet in {\em Xin Tang shu} 32.838 (Ouyang 1975).  
The preceding day, the 17th day of the 
lunar
month corresponding to the 35th day 
in the sexagenary calendar ({\em wuxu}), AD 767 Jan 21, 
is given for the same event in {\em Tang hui yao} 43.4 (Wang 2007).  

In his notes on comet no. 437, Hasegawa (1980) cites an unspecified Chinese record, 
which appears to include the portion of the {\em Gujin tushu jicheng} record pertaining to the comet, 
Hasegawa gives a date, which would correspond to AD 776 Jan 11,
but is based on an erroneous {\em yihai} date.
With the exception of the date, 
the unspecified record corresponds very closely with descriptions of the AD 767 Jan 21/22 comet.

{\it Xin Tang shu} 32.838 (Ouyang 1975):
\begin{quotation}
On a jihai day [day 36 of the sexagenary cycle] in the 12th month 
of the first year of Dali reign-period (AD 767 Jan 22), a comet appeared in Hu Gua. 
It was greater than one chi in length. 
After 20 days it was no longer seen. It fell into Huanzhe.
\end{quotation}
{\it Tang hui yao} 43.4 (Wang 2007) has a similar entry corresponding to
Hasegawa (1980) for comet no. 433:
\begin{quotation}
On the 17th day [of the lunar month, 
corresponding to the 35th day of the sexagenary cycle, 
{\em wuxu}] of the 12th month of the first year of the Dali reign period 
(AD 767 Jan 21\footnote{According to eclipse.gsfc.nasa.gov/phase, new moon was AD 767 Jan 5 at 8:13h UT,
i.e. around 15:30h local time in China,
the 1st day of that month, so that the {\em 17th} day of that lunar month was Jan 21. 
The 17th night of that month was Jan 21/22.
The Chinese started the day-count
in each month with what we call {\em new moon}, i.e. conjunction of moon and sun,
as confirmed by the fact that all dates of solar eclipses from AD 700 to 1200
are dated to {\em the first day of the month}, see listing in Xu et al. (2000).}) 
a comet appeared in Hu Gua. It encroached on Huanzhe.  
It was greater than one chi in length, and its colour was red.  
After 20 days it vanished.
\end{quotation}
{\it Jiu Tang shu} 36.18 (Liu 1975): 
\begin{quotation}
In the 12th month of the first year of the Dali reign period 
a comet appeared in Hu Gua. It was greater than one chi in length. It fell into Huanzhe.
\end{quotation}

AD 767 Jan 22 was a {\em jihai} day.  
The initial appearance of the comet was in the {\it Xin Tang shu} recorded 
as having occurred on Jan 22. 
It would be very easy for {\em ji} to be mistranscribed as the very similar graph {\em yi},
whether due to hastiness on the part of a scribe, 
or due to reproduction of a damaged manuscript, 
hence changing a {\em jihai} day to an {\em yihai} day. 
As noted in Yau et al. (1995), this particular error is extremely common.  
In AD 776 Jan, there was an {\em yihai} day.
It is also quite conceivable that {\em yuan} might be mistranscribed as {\it shi}, 
hence changing the date from the 1st year of the Dali reign-period, AD 767, 
to the 10th year, AD 776.  
Given the remarkable similarities between the AD 767 comet and the presumable AD 776 comet, 
including the period for which they were visible,
their lengths, and the particular constellations in which they appeared and toward which they traveled,
we find it very likely that they were indeed the same comet 
and that the AD 776 date is the result of some error.  
The presumable AD 776 comet is also not listed in Ho Peng Yoke (1962).

What is reported by Chen et al. (1934) as
{\em white qi above the moon penetrated Zui, Shen, Jing (Gem), Gui (Cnc), Liu, and Xuanyuan},
was neither identified nor considered as an aurora in any of the catalogues cited above.
Indeed, it was not an aurora,
because it appeared {\em above the moon}, similar to the event in Sect. 2.7.
Instead, it could be the same event:
While the phrases {\em in the east}, {\em ten bands}, Taurus and Auriga, 
and the {\em dispersing} were not mentioned in the Chen et al. (1934) source,
the constellations mentioned to be above the moon are fully consistent with 
a dating around the full moon -- given that the {\em white qi} phenomenon occurred in Jan 776 
(Chen et al. 1934: {\em in the 12th month}). 
It could have happened on the {\em bingzi day} as mentioned in Sect. 2.7,
i.e. it may have been the same event.
\footnote{While the comet was misdated to AD 776 Jan, 
we have to consider whether the {\em white qi} from the late {\em Gujin tushu jicheng} source 
(which gives year and month without day) may also be dated to AD 767 Jan.
The event discussed in Sect. 2.7 gives the
{\em bingzi day of the 12th month [AD 776 Jan 12] (the moon rose in the east
and above it there were more than 10 bands of white qi)}.
There was no {\em bingzi day} (13th day of the sexagenary cycle)
in the 12th month
of that year in the Chinese lunar calendar, which started on AD 767 Jan 5 with the new moon.
Hence, the event discussed in Sect. 2.7 could not have happened in AD 767. \\
We still have to consider, whether the {\em yihai} day from that unspecified source
(reporting only the comet)
may be related to the {\em white qi} on a {\em bingzi} day (see Sect. 2.7).
The {\em white qi} could have happened on AD 776 Jan 11
and continued after midnight into the {\em bingzi} day AD 776 Jan 12 (see footnote 6).}

The {\em Gujin tushu jisheng} (without specifying a day) appears, therefore, 
to have conflated two separate events -- {\em white qi} and comet -- 
on account of their (possibly) similar dates within the traditional calendar.
The white {\em qi} phenomenon most probably occurred 
on the night of AD 776 Jan 11/12 
and/or 12/13 (see Sect. 2.7,
halos sometimes do appear in subsequent nights), 
while the comet appeared on AD 767 Jan 21/22.

We also see that the Chinese reports -- after correction of some mistakes
in dates due to likely transcription erors -- are fully consistent in all details,
so that the date correction is unambiguous and highly credible.

\subsection{The possible Korean comet in AD 776 June}

From the {\em Samguk sagi}, the Korean Lee Dynasty chronology during the reign of He Gong 
Chu (1968) cited a {\em Hye Sung} interpreted as nova or supernova candidate,
which was observed from AD 776 Jun 1 to 30 in Tau-Aur.

Unfortunately, as Chu (1968) does not give a chapter or page in his citation of the {\em Samguk sagi},
we are unable to locate the event to which Chu (1968) refers in the historical chronicle.
We might expect to see a record in the {\em Basic Annals} for the 12th year of the reign of King Hyegong 
(AD 776), but no such record is present.  
The {\em Samguk sagi} contains neither an omenological nor an astronomical treatise, 
so we can only assume that Chu (1968) refers to a passage contained 
in either a biographical chapter or another treatise in the text.
The next closest comet in time around AD 776 
is an AD 768 event reported in the {\em Silla pon'gi} section of 
{\em Samguk sagi} (9.95): 
\begin{quotation}
In the fourth year [of the reign of King Hyegong] in spring, a comet appeared in the NE.
\end{quotation}
(Silla pon'gi, Samguk sagi 9.95 / Kim 1977).

Since {\em Hye Sung} is an ancient Korean comet name,
the object may be a comet observed for 30 days
(it may have been referred to by a name for a comet, because it moved relative to the stars).
The fact that Chu (1968) interpreted it as a nova or supernova may be due to
the fact that 
neither a tail nor motion
was mentioned in the original text.
If it was a comet with a very short or faint (undetectable) tail,
then this may indicate low solar activity at that time (June 776).
The fact that it was not reported in any other source
may also indicate that it was relatively faint.

\section{Conclusion: A cluster of East Asian aurora reports in the 770s~?}

Based on new translations, philological considerations, historic backgrounds,
precise date and time conversions, and astronomical calculations, 
we have shown that several historic Chinese reports were misinterpreted as aurorae, 
and that they were in fact something completely different, 
e.g. solar or lunar halos, fog, meteors, and so forth.
In particular, misinterpretations and errors in dating 
call into question the conclusions of both Usoskin et al. (2013) and Zhou et al. (2014).

We conclude that there is neither evidence for a cluster of aurorae nor for any particularly strong 
aurorae in the AD 770s in East Asian sources.
However, the lack of aurora reports for the AD 770s is not due to the absence of observers or the loss of records,
as there are Chinese reports about (other) celestial events from the decade.
In particular, there are no aurora reports which would support the super-flare hypothesis for AD 774.

There is a distinct cluster of very likely aurorae in AD 762;
and there are good candidates in AD 757, 760, 767, and 770 (one or two), 
which are, however, questionable for a variety of reasons outlined in section 4.

Stephenson (2015) cites the events included in our Sect.
3 also as good examples of true aurorae. Of the eight
events listed by us from AD 767 to 779, he lists six (all
excluding those in our Sect. 2.4 and 2.6) plus one more
({\em black qi}). Stephenson (2015)
does not consider the {\em white qi} observation in connection
with the Chinese comet observation, the latter misdated for AD 776
(true AD 767). Stephenson (2015) does classify the event
of AD 776 Jan 12 {\em above the moon}
as an aurora, which we consider to be a
halo effect.
Stephenson (2015) and we agree that there is no evidence for strong
East Asian aurorae around AD 774/5.
As indicated by Stephenson (2014), the records of astronomical observations dating to the Dali reign period 
(AD 767-779) are particularly numerous. It is unlikely that any major observable astronomical phenomenon 
that occurred during that period is missing from its records.

As far as European observations of presumable aurorae in the mid AD 770s are concerned, 
as listed e.g. in Usoskin et al. (2013) or the Silverman catalog, 
for AD 773, 774, and 776, Neuh\"auser \& Neuh\"auser (2014, 2015) 
have shown that all of them were in fact halo displays;
two of those are also
listed in Stephenson (2015), who considers at least one of them,
the {\em red cross} of AD 776, as a possible aurora.

The two Chinese comet reports in the mid AD 770s also do not relate to the $^{14}$C increase: 
The comet of Jan 773 did not collide with the Earth's atmosphere (Chapman et al. 2014), 
and the comet reported for AD 776 was actually observed nine years earlier (Sect. 5.2).

Comet tail length and, hence, brightness, visibility, duration, 
and detection rates of comets depend on the solar wind. 

For AD 773 Jan 17, there is a report about a Chinese comet: 
{\em ... there was a long star beneath Shen (Orion). Its length extended across the sky ...}.
See Sect. 5.1 and Chapman et al. (2014).
This comet was also observed by the Japanese on (or since) AD 773 Jan 20; 
the fact that the observations are reported only for a few days (Jan 17 and 20) 
does not mean that it was observed solely in those nights, but could mean that it was
observed first on those dates (Chapman et al. 2014). 
Stephenson (2015) also {\em reexamines} the evidence
concerning the AD 773 comet, corroborating the conclusions of Chapman et al. (2014).
Moreover, Hase\-ga\-wa (1980) and Kronk (1999) list a Chinese comet for AD 776 Jan 11, 
but that is a misdated copy of the comet observed in AD 767
(Ho Peng Yoke (1962) does not list this extra comet),
see Sect. 5.2. 

Next, the nova or supernova candidate listed as {\em Hye Sung} 
in Chu (1968) for AD 776 Jun 1-30 in Tau-Aur 
(from the Korean Lee Dynasty chronology during the reign of He Gong) 
may be a comet, 
because it was named with an old Korean comet name 
(see Sect. 5.3).
This event is not examined in Stephenson (2015).

According to Schove (1984), Ho Peng Yoke (1962), Hase\-ga\-wa (1980), and Kronk (1999),
there were no (other) comets detected worldwide until at least AD 812. 
This could be consistent with relatively strong solar wind (and related aurorae) until at least AD 773 Jan,
but weak solar wind (few or no aurorae and short or no comet tails) 
since at least AD 776 June
(for aurorae, see Neuh\"auser \& Neuh\"auser 2015).
This consideration is speculative, because one would need to know the distance of the comet
to convert from apparent to true (projected) length -- but an apparent comet tail length might
be considered a good first approximation.

\acknowledgements
We obtained the moon phases from Fred Espenak, NASA/GSFC (eclipse.gsfc.nasa.gov) and
from Rita Gautschy on www.gautschy.ch/$\sim$rita/archast/mond/Babylonerste.txt.
We consulted the au\-ro\-ra catalog of Silverman on 
nssdcftp.gsfc.nasa.gov/miscellaneous/aurora.
We thank V. Hambaryan for pointing us to the Korean comet observation of AD 776.

{}

\end{document}